\documentclass[11pt,a4paper,openright]{article}

\usepackage{amsmath}

\usepackage{natbib}
\usepackage{booktabs}
\usepackage{graphicx}

\usepackage{times}
\usepackage[varg]{txfonts}

\usepackage{graphicx}
\usepackage{balance}  
\usepackage{color}
\usepackage{algorithm}
\usepackage{algorithmic}
\usepackage[affil-it]{authblk}
\usepackage{authblk}

\begin{document}

\newtheorem{theorem}{Theorem}
\newtheorem{corollary}[theorem]{Corollary}
\newtheorem{lemma}{Lemma}
\newtheorem{observation}[theorem]{Observation}
\newtheorem{proposition}[theorem]{Proposition}
\newtheorem{definition}{Definition}
\newtheorem{problem}{Problem}
\newtheorem{claim}[theorem]{Claim}
\newtheorem{fact}[theorem]{Fact}
\newtheorem{assumption}[theorem]{Assumption}
\newtheorem{example}{Example}
\newtheorem{question}{Question}

\newenvironment{proof-sketch}{\noindent{\bf Sketch of Proof:}\hspace*{1em}}{\qed\bigskip}
\newenvironment{proof-idea}{\noindent{\bf Proof Idea:}\hspace*{1em}}{\qed\bigskip}
\newenvironment{proof-of-lemma}[1]{\noindent{\bf Proof of Lemma #1:}\hspace*{1em}}{\qed\bigskip}
\newenvironment{proof-attempt}{\noindent{\bf Proof Attempt:}\hspace*{1em}}{\qed\bigskip}
\newenvironment{proofof}[1]{\noindent{\bf Proof} of #1:\hspace*{1em}}{\qed\bigskip}
\newenvironment{remark}{\noindent{\bf Remark}\hspace*{1em}}{\bigskip}
\newcommand{\tab}{\hspace*{2em}}
\newcommand{\mbf}[1]{\mbox{{\bfseries #1}}}
\newcommand{\N}{\mbf{N}}
\newcommand{\<}{\langle}
\renewcommand{\>}{\rangle}
\definecolor{mycolor}{rgb}{0.0,0.0,0}
\mathchardef\mhyphen="2D

\newcommand{\eat}[1]{} 
\newcommand{\ratone}{\mbox {$\rho_1$}} 
\newcommand{\rattwo}{\mbox {$\rho_2$}} 
\newcommand{\bwp}{\mbox {\tt BWP}} 
\newcommand{\ab}{\mbox {\tt AB}} 
\newcommand{\calr}{\mbox {${\cal R}$}} 
\newcommand{\cals}{\mbox {${\cal S}$}} 
\newcommand{\unai}{\mbox {$U_i$}} 
 \newcommand{\unaone}{\mbox {$U_1$}} 
\newcommand{\unatwo}{\mbox {$U_2$}} 
\newcommand{\unathree}{\mbox {$U_3$}} 
\newcommand{\unafour}{\mbox {$U_4$}} 
\newcommand{\unafive}{\mbox {$U_5$}} 
\newcommand{\unam}{\mbox {$U_m$}} 
\newcommand{\unaM}{\mbox {$U_M$}} 
\newcommand{\polonefive}{\mbox {$P_{1,5}$}} 
\newcommand{\pol}{\mbox {$P_{1,M}$}} 
\newcommand{\rjup}{\mbox {$r_j^{+}$}} 
\newcommand{\rjdown}{\mbox {$r_j^{-}$}} 
\newcommand{\rup}[1]{\mbox {$r_{#1}^{+}$}} 
\newcommand{\rdown}[1]{\mbox {$r_{#1}^{-}$}}  

\newcommand{\fopt}{\mbox {$F^{opt}$}}

\newcommand{\LL}[1]{\textcolor{black}{#1}} 
\newcommand{\NK}[1]{\textcolor{blue}{#1}} 
\newcommand{\DD}[1]{\textcolor{red}{#1}} 
\newcommand{\LLrev}[1]{\textcolor{green}{#1}} 
\newcommand{\KN}[1]{{{\bf {#1}}}}

\title{Querying Temporal Drifts at Multiple Granularities}

\date{}

\author{Sofia Kleisarchaki}
\author{Sihem Amer-Yahia} 
\affil{University Grenoble Alps, CNRS, France}
\author{Ahlame Douzal Chouakria}
\author{Vassilis Christophides}
\affil{CSD, UoC, Greece \& Inria, Paris-Rocquencourt, France}
\author{Ruben H. Zamar}
\affil{Univ. of British Columbia, Canada}

\maketitle

\begin{abstract}
There exists a large body of work on online drift detection with the
goal of dynamically finding and maintaining changes in data
streams. In this paper, we adopt a query-based approach to drift
detection. Our approach relies on {\em a drift index}, a structure
that captures drift at different time granularities and enables
flexible {\em drift queries}. We formalize different drift queries
that represent real-world scenarios and develop query evaluation
algorithms that use different materializations of the drift index as
well as strategies for online index maintenance. We describe a
thorough study of the performance of our algorithms on 
real-world and synthetic datasets with varying change rates.
\end{abstract}

\section{Introduction}

Monitoring streaming content is a challenging  big data analytics 
problem, given that very large datasets are rarely (if ever) stationary. 
In several real world monitoring applications (e.g., newsgroup 
discussions, network connections, etc.) we need to detect significant 
change points in the underlying data distribution (e.g., frequency of 
words, sessions, etc.) and track the evolution of those changes over 
time. These change points, depending on the research community, are 
referred to as {\em temporal evolution}, {\em non
  stationarity}, or {\em concept drift} and provide
valuable insights on real world events (e.g. a discussion
topic, an intrusion) to take a timely action. 
In this paper, we adopt {\em a query-based approach to drift detection}
and address the question of processing {\em drift queries} over very
large datasets. To the best of our knowledge, our work is the first to
formalize flexible drift queries on streaming datasets with varying
change rates.

In the problem of drift detection, given a number of $m$ drifts ordered in time, we need no less than $m+1$ intervals to detect them. Thus, without any assumption on the underlying distribution, we are interested in exploring how to segment the input stream in order to find a reasonable tradeoff between true positives and false negatives. Existing methods rely on segmenting the input stream, mostly into smaller fixed length intervals [3, 5, 9, 11, 14]. Although some works exist on partitioning the same stream into intervals of different granularities [2, 8, 15] they either adopt an offline analysis or they lack the ability of querying historical drifts in streams. 

A granularity in this case is an interval of time (e.g., every hour)
or a number of observed data points (e.g., every 200 points). A drift is then defined as a significant
difference in data distributions between two consecutive intervals
at the same granularity. To detect drifts either statistical tests are directly applied on the data of two intervals [6, 11] or on their summaries, as for instance provided by a clustering algorithm [1, 3, 4]. 
To this end, two parameters impact
the accuracy and efficiency of drift detection: the granularity of the
intervals at which the original data items are clustered and the drift
significance threshold used to assess whether or not there is a drift
between two consecutive clusterings. In fact, fine-grained
intervals can be used to capture the evolution of frequently changing
streams. However, they may induce computation overhead for slowly changing ones. In
addition, they may cause false positives, i.e., detecting drifts that
are too sudden and noisy, hence hurting precision. While a coarser
granularity will improve precision, since more data is clustered in
each interval, it may incur missing a drift that occurred at a finer
granularity. Those misses will negatively affect recall. Moreover,
the rate of change of a given dataset may vary over time thereby
requiring to consider different clustering granularities and drift
thresholds for the same dataset.

Understanding the tradeoff between precision (at higher segmentation granularities)
and recall (at lower segmentation granularities), and the choice of
thresholds to determine what constitutes a drift between two
consecutive intervals of the same granularity, are the main objectives
of this work. In this paper, we adopt an analytics approach in which
we formalize drift queries over both fresh and historical data of
arbitrary time granularities, in order to provide flexibility in
tracking and analyzing drifts in evolving datasets. 
For this reason,
we propose a flexible drift index to organize past data (or more
precisely their summaries) at several granularities. Furthermore. we explore different creation strategies for this index relying on two common clustering approaches, namely independent [7, 12, 16] and cumulative [1, 4]. In independent clustering, data points belonging to a given interval are considered equally important and clustered independently. In cumulative clustering, data points in a given interval are clustered with all previously occurring points and fresher data is more important than older data. Moreover, we propose different materialization strategies in order to explore the tradeoff between index storage and query response time. 

Unlike existing approaches [3, 5, 9, 11, 14] comparing only the last  most recent intervals, we exploit this index in order to identify drifts at different granularities. In particular, we formalize three kinds of queries: unary, refinement and synthesis aiming to detect drifts against historical data. 
A unary query is used to extract all drifts detected at
a given granularity. A refinement query explores drifts from a source
granularity (e.g., 5,000 points) to a finer target granularity (e.g.,
500 points), iteratively. Such a query is useful to provide a more
detailed description of drifts that have been detected in a high
granularity, resulting in better recall. Synthesis queries, on
the other hand, start from a relatively low granularity and summarize
them into coarser ones. In this case, some of the particular details
might be missed (low recall) in order to get drifts with higher
precision. 
This flexibility in querying drifts allows us to explore, in a declarative fashion,
precision and recall tradeoffs at different granularities. Also, it addresses a long
standing concern in detecting and tracking drifts in streaming
content, namely adaptability of drift detection to different drift arrival rates and types.

The evaluation of declarative drift queries relies on traversing the 
index of historical data summaries and, at each granularity, comparing 
its nodes pairwise to identify points where clusterings 
dissimilarity exceeds a threshold $\theta$. Rather than setting drift 
thresholds a-priori [5, 13], we learn a $\theta$-value for each dataset 
and at each granularity level in the index. 

In summary, this paper makes the following contributions: 
\begin{enumerate}
\itemsep-0.4em
\item We introduce and formalize drift queries that provide high
  flexibility in analyzing precision and recall of drift
  detection for different time granularities.
\item We propose a drift index, a graph structure that captures change
  at different granularities and explore different materializations of
  the index that lead to the design of various index maintenance and
  query evaluation algorithms.
\item We propose learning algorithms for learning drift and clustering thresholds adaptively for different granularities and rates of change.
\item We perform a thorough study of proposed queries and indices using two real datasets, KDD Cup'99~\footnote{\small \em https://archive.ics.uci.edu/ml/datasets/KDD+Cup+1999+Data} 
and Usenet [10], and a synthetically generated dataset. On the effectiveness front, our study confirms the need for our refinement and synthesis queries, as demonstrated by the very good precision/recall results they attain. On the scalability front, it validates the need for different materializations of the drift index in order to achieve a tradeoff between storage and query response time for datasets of varying change rates.
\end{enumerate}

\section{Drift Queries} 
\label{sec:drift-queries}
The goal of drift queries is to compare drifts at different
granularities and provide analysts with the ability to explore drift
precison and recall across granularities. We study two kinds of
queries, {\em refinement} and {\em synthesis}. Both kinds rely on a
simpler {\em unary query} defined as follows.

\begin{definition}
{\bf Unary Query.} A unary query $UQ(D,g)$ returns the set of all drifts $X^g$ detected
at granularity $g$ for a dataset $D$.
\end{definition}

\begin{definition}
{\bf Refinement Query.} A refinement query $RQ(D,g_s,g_t)$ admits a source granularity $g_s$
and a target one $g_t$ s.t. $g_t \prec\prec g_s$, and returns a set of
pairs $(x_i^{g_s},x_j^g)$ where each drift $x_i^{g_s} \in X^{g_s}$ at
$g_s$ is associated to the finest corresponding drift
$x_j^g \in X^g$ at a granularity $g$ no finer than $g_t$ as follows:
\end{definition}

\begin{math}
\{x_j^g \in X^g, g_t \prec\prec g \prec\prec g_s \vee g=g_t \ | \\
\exists x_i^{g_s} \in X^{g_s}, I_j^g \subseteq (I_{i}^{g_s} \cup I_{i+1}^{g_s}), \\
\nexists x_k^{g'} \in X^{g'}, g_t \prec\prec g' \prec\prec g \vee g'=g_t, I_k^{g'} \subseteq (I_{i}^{g_s} \cup
I_{i+1}^{g_s})\}
\end{math} \\

where $I_{i}^{g_s} \cup I_{i+1}^{g_s} = [\displaystyle\min_{ts_j \in I_i^{g_s}} (ts_j), \max_{ts_k \in I_{i+1}^{g_s}} (ts_k)]$ and \\
$I_j^g \subseteq (I_{i}^{g_s} \cup I_{i+1}^{g_s})$ if $\displaystyle\min_{ts_j \in (I_i^{g_s} \cup I_{i+1}^{g_s})} (ts_j) \leq \min_{ts_k \in I_i^{g_s}} (ts_k)$ and $\displaystyle\max_{ts_k \in I_i^{g_s}} (ts_k) \leq \max_{ts_j \in (I_i^{g_s} \cup I_{i+1}^{g_s})} (ts_j)$\\

Refinement queries provide a detailed analysis of drifts iteratively. For instance, for a source granularity $g_s=1000$
connections on KDD Cup'99, selecting a granularity
$g_t=500$ might result in missing a more insightful analysis occurring
at granularity $g_t=100$. On the other hand, selecting $g_t=100$ may
result in retrieving false positives which could be avoided at
$g_t=500$. Therefore, the analyst will use the refinement query
$RQ(D,1000,100)$ to obtain details of each drift at $g_s=1000$ with a
tradeoff between false negatives and false positives.

\begin{definition}
{\bf Synthesis Query.} A synthesis query\\ $SQ(D,g_s,g_t)$ admits a source granularity $g_s$
and a target one $g_t$ s.t. $g_s \prec\prec g_t$, and returns a set of
pairs $(x_i^{g_s},x_j^g)$ where each drift $x_i^{g_s} \in X^{g_s}$ at
granularity $g_s$ is associated to the coarsest corresponding drift
$x_j^g \in X^g$ at a granularity $g$ no coarser than $g_t$ as follows:
\end{definition}

\begin{math}
\{x_j^g \in X^g, g_s \prec\prec g \prec\prec g_t \vee g=g_s \ | \\
\ \exists x_i^{g_s} \in X^{g_s}, I_i^{g_s} \subseteq (I_{j}^g \cup I_{j+1}^g), \\
\nexists x_k^{g'} \in X^{g'}, g \prec\prec g'
\prec\prec g_t \vee g'=g_t, I_i^{g_s} \subseteq (I_{k}^{g'} \cup
I_{k+1}^{g'})\}
\end{math}
\ \\ 

Synthesis queries provide a summary analysis of drifts iteratively. For instance, for a source granularity $g_s=100$ connections
on KDD Cup'99, selecting a granularity $g_t=1000$ might
result in missing a more precise synthesis occurring at
$g_t=2000$. On the other hand, selecting $g_t=2000$ can
result in missing a summary of a drift, which could be obtained at
$g_t=1000$. Therefore, the analyst can use the synthesis
query $SQ(D,100,2000)$ to obtain a summary of each drift at $g_s=100$
with a tradeoff between false negatives and false positives.

\section{References}

{\setlength{\parindent}{0cm}
[1] C. C. Aggarwal, J. Han, J. Wang, and P. S. Yu. A framework for clustering evolving data streams. In Proc. of VLDB’03 - Volume 29, pages 81–92.

[2] A. Bifet. Adaptive stream mining: Pattern learning and mining from evolving data streams. In Proc. of the 2010 Conference on Adaptive Stream Mining:
Pattern Learning and Mining from Evolving Data Streams, pages 1–212, Am- sterdam, The Netherlands, 2010. IOS Press.

[3] A. Bondu, B. Grossin, and M.-L. Picard. Density estimation on data streams : an application to change detection. In EGC’10, pages 229–240.

[4] F. Cao, M. Ester, W. Qian, and A. Zhou. Density-based clustering over an evolving data stream with noise. In SIAM CDM’06, pages 328–339.

[5] T. Dasu, S. Krishnan, S. Venkatasubramanian, and K. Yi. An information- theoretic approach to detecting changes in multi-dimensional data streams. In Proc. Symp. on the Interface of Statistics, Computing Science, and Appli- cations, 2006.

[6] A. Dries and U. Rckert. Adaptive concept drift detection. Statistical Analysis and Data Mining, 2(5-6):311–327, 2009.

[7] M. Ester, H.-P. Kriegel, J. S, and X. Xu. A density-based algorithm for dis- covering clusters in large spatial databases with noise. pages 226–231. AAAI Press, 1996.

[8] J. Gama, P. Medas, G. Castillo, and P. Rodrigues. Learning with drift detec- tion. In Advances in AI – SBIA’04, volume 3171 of Lecture Notes in Computer Science, pages 286–295. Springer Berlin Heidelberg.

[9] N. Japkowicz, C. Myers, and M. Gluck. A novelty detection approach to classification. In Proc. of AI, pages 518–523, 1995.

[10] I. Katakis, G. Tsoumakas, and I. Vlahavas. Dynamic feature space and in- cremental feature selection for the classification of textual data streams. In in ECML/PKDD-2006, page 107. Springer Verlag.

[11] D. Kifer, S. Ben-David, and J. Gehrke. Detecting change in data streams. In Proc. of VLDB’04 - Volume 30, pages 180–191.

[12] L. O’Callaghan, N. Mishra, M. A., S. Guha, and R. Motwani. Streaming-data algorithms for high-quality clustering, 2001.

[13] R. Sebastia ̃o and J. a. Gama. Change detection in learning histograms from data streams.

[14] P. Vorburger and A. Bernstein. Entropy-based concept shift detection. In ICDM’06, pages 1113–1118. IEEE Computer Society.

[15] G. Widmer and M. Kubat. Learning in the presence of concept drift and hidden contexts. In ML, pages 69–101, 1996.

[16] T. Zhang, R. Ramakrishnan, and M. Livny. Birch: an efficient data clustering method for very large databases. SIGMOD Rec., 25:103–114, 1996.
}
\end{document}